\documentclass[aps,preprint,superscriptaddress,floatfix]{revtex4-1}

\usepackage{mathptmx}
\usepackage[scaled=0.9]{helvet}
\usepackage[utf8]{inputenc}
\usepackage{graphicx}
\usepackage[dvipsnames]{xcolor}
\usepackage{textcomp}
\usepackage{ifsym}
\usepackage{mathrsfs}
\usepackage{sidecap}
\usepackage{xspace}

\newcommand{\celsius}{$^{\circ}$C\xspace}

\begin{document}

\title{N-polar GaN \emph{p-n} junction diodes with low ideality factors}

\author{Kazuki~Nomoto}
\affiliation{School of Electrical and Computer Engineering, Cornell University, Ithaca, New York 14853, USA}
\author{Huili~Grace~Xing}
\affiliation{School of Electrical and Computer Engineering, Cornell University, Ithaca, New York 14853, USA}
\affiliation{Department of Materials Science and Engineering and Kavli Institute for Nanoscale Science, Cornell University, Ithaca, New York 14853, USA}
\author{Debdeep~Jena}
\affiliation{School of Electrical and Computer Engineering, Cornell University, Ithaca, New York 14853, USA}
\affiliation{Department of Materials Science and Engineering and Kavli Institute for Nanoscale Science, Cornell University, Ithaca, New York 14853, USA}
\author{YongJin~Cho}
\email[Author to whom correspondence should be addressed: ]{yongjin.cho@cornell.edu}
\affiliation{School of Electrical and Computer Engineering, Cornell University, Ithaca, New York 14853, USA}

\begin{abstract}
High-quality N-polar GaN \emph{p-n} diodes are realized on single-crystal N-polar GaN bulk substrate by plasma-assisted molecular beam epitaxy. The room-temperature current-voltage characteristics reveal a high on/off current ratio of $>10^{11}$ at $\pm4$~V and an ideality factor of 1.6. As the temperature increases to 200~\celsius, the apparent ideality factor gradually approaches 2. At such high temperatures, Shockley-Read-Hall recombination times of 0.32--0.46~ns are estimated. The measured electroluminescence spectrum is dominated by a strong near-band edge emission, while deep level and acceptor-related luminescence is greatly suppressed. A relatively high reverse breakdown field of 2.4~MV/cm without field-plates is achieved. This work indicates that the quality of N-polar GaN diodes is now approaching to that of their state-of-the-art Ga-polar counterparts.  

\end{abstract}
\maketitle

In noncentrosymmetric wurtzite III-nitride heterostructures, spontaneous and piezoelectric polarization fields on the order of MV/cm (Ref.~\onlinecite{bernardini1997spontaneous}) along the polar \emph{c}-axis can generate 2D electron and hole gases,\cite{ambacher1999two,cho2019high,chaudhuri2019polarization,zhang2021polarization} facilitate the activation of deep acceptors,\cite{simon2010polarization,li2013polarization} and even create doping effects without chemical dopants\cite{li2012polarization}.
Furthermore, such large built-in polarization
fields can be engineered for innovative applications in photonic and electronic devices.\cite{jena2011polarization,li2015polarization,wood2007polarization,cho2019blue} In this regard, the N-polar direction is as important as the metal-polar one and is advantageous for its own unique applications in devices such as buried-barrier high electron mobility transistors,\cite{wong2013n,wienecke2016n} interband tunnel junctions\cite{krishnamoorthy2010polarization,yan2015polarization} and resonant tunneling diodes\cite{cho2020n}.

From the perspective of the molecular beam epitaxy (MBE) growth of GaN, the two surface polarities behave quite differently: the contracted Ga bilayer\cite{northrup2000structure}, which is beneficial for smooth surface morphology in Ga-polar GaN growth,\cite{northrup2000structure,koblmuller2005ga} is \emph{unstable} on the N-polar surface.\cite{monroy2004growth} Instead, the Ga adlayer on the N-polar surface is less than 1~monolayer thick before Ga droplet formation begins,\cite{monroy2004growth} meaning that growth with smooth surface morphology is relatively harder for the N-polar surface. It is also known that under a fixed growth condition, the N-polar growth exhibits higher O incorporation and much lower Mg incorporation, by almost an order of magnitude, compared to metal-polar surface growth,\cite{ptak2001controlled,sumiya2000dependence,ptak2001magnesium} which can make \emph{p}-doping of N-polar GaN difficult.     

Despite these challenges, there has been progress on N-polar GaN homoepitaxy,\cite{okumura2014growth,turski2019unusual,wurm2020growth} owing to the recent commercial availability of high-quality N-polar bulk GaN substrates, as well as N-polar device fabrication such as \emph{p-n} diodes\cite{cho2017single} and resonant tunneling diodes\cite{cho2020n}. Although we recently reported the successful growth of N-polar single-crystal GaN \emph{p-n} diodes with high structural quality and a high diode on/off current ratio on GaN bulk crystal, the
quality of the devices was, however, far from that of the state-of-the-art Ga-polar diodes.\cite{cho2017single,hu2015near} Specifically, high-quality N-polar GaN \emph{p-n} diodes with an ideality factor less than 2 have not been demonstrated yet.      

In this paper, we report the temperature dependent current-voltage characteristics, electroluminescence, and the Shockley-Read-Hall (SRH) recombination time of single-crystal N-polar GaN \emph{p-n} junction diodes with low ideality factors, which is enabled by combining high-quality N-polar GaN bulk substrates and optimized growth conditions for doping.

\begin{figure}[t!]
\centering
\includegraphics*[width=8cm]{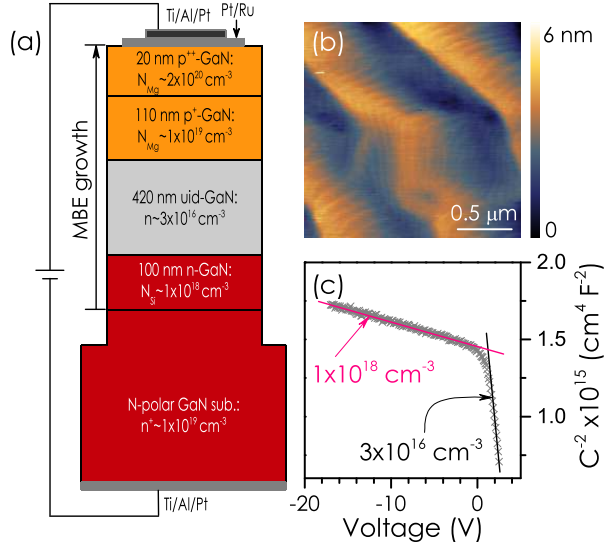} 
\caption{(a) Schematic layer structure of N-polar GaN \emph{p-n} diode devices. (b) $10\times10$~$\mu$m$^{2}$ AFM micrograph of the as-grown sample surface. (c) \emph{C}$^{-2}$ versus voltage plot using measured junction capacitance \emph{C}. Two linear fittings extract the net doping concentrations in the unintentionally doped GaN (black line) and the 100~nm Si-doped GaN buffer layer (pink line).}
\label{structure}
\end{figure}

N-polar GaN \emph{p-n} diode structures were grown on single-crystal N-polar $n^{+}$-type GaN$(000\bar{1})$ bulk substrate\cite{zajac2018basic} with a dislocation density of $5\times10^{4}$~cm$^{-2}$ and a mobile electron density of $\sim$10$^{19}$~cm$^{-3}$ in a Veeco GENxplor MBE reactor equipped with standard effusion cells for elemental Ga, Si, Mg and a radio-frequency plasma source for the active N species. Si and Mg were used as the \textit{n}-type and \textit{p}-type dopant, respectively. The base pressure of the growth chamber was in the range of 10$^{-10}$~Torr under idle conditions, and $1.5\times10^{-5}$~Torr during growth. Starting from the nucleation substrate surface, the N-polar GaN \emph{p-n} diode structure consists of 100~nm of GaN:Si followed by 420~nm unintentionally doped (uid) GaN, 110~nm GaN:Mg, then 20~nm GaN:Mg. The details of the layer structures are shown in Fig.~\ref{structure}(a). The top heavily doped GaN:Mg layer is intended for a low resistance \emph{p}-contact. All of the layers were grown under Ga-rich conditions in order to guarantee smooth morphology and minimal O impurity incorporation. The Si and Mg dopant densities used in the doped layers have been estimated from separate secondary ion mass spectrometry (SIMS) stack samples. \emph{In situ} reflection high energy electron diffraction (RHEED) revealed that the MBE-grown layers showed a clear $(3\times3)$ surface reconstruction below 200~\celsius after growth, confirming the N-polarity of the diode structure.\cite{smith1998reconstructions,cho2017single} The excess Ga droplets after the growth were removed in HCl before any \emph{ex situ} characterization and device fabrication. Atomic force microscopy exhibits smooth surface morphology with a root-mean-square roughness of 0.6~nm on a $2\times2$~$\mu$m$^{2}$ [Fig.~\ref{structure}(b)].   


The devices were fabricated using a general mesa-design by a plasma etching without applying any edge termination. In order to fabricate the \emph{p-n} junction diodes, the epitaxially grown sample was first cleaned by solvents, HF and HCl. The mesa structure was fabricated by inductively-coupled plasma (ICP) dry-etching, after which, circular alloyed Pt/Ru anode ohmic electrodes and Ti/Al/Pt thick metal pads were formed by a lift-off process on the surface of the \emph{p}$^{++}$-GaN cap layer. The Pt/Ru ohmic contacts were alloyed at 650~\celsius in O$_{2}$ ambient. A Ti-based cathode ohmic electrode was formed on the rear surface of the GaN substrate.

In order to estimate the doping concentrations in the
uid-GaN region, a capacitance versus voltage (\emph{C-V}) measurement was performed on the N-polar GaN \emph{p-n} diode at a frequency of 1~MHz. Loss tangent angles $>89^{\circ}$ indicate very low leakage and high reliability of the measurement. The corresponding $C^{-2}$-$V$ plot in Fig.~\ref{structure}(c) exhibits two distinct slopes. For one-side abrupt junctions, as in the case of the \emph{p-n} diodes in this study, $dC^{-2}/dV \propto -N^{-1}$, where $N$ is the dopant density of the lightly doped side.\cite{sze2007physics} Thus, the observed two different slopes correspond to the net donor densities in the uid-GaN and the Si-doped GaN nucleation layer [Fig.~\ref{structure}(c)]. This net donor density of $3\times 10^{16}$~cm$^{-3}$ in the uid-GaN layer is remarkably low for N-polar GaN in that O impurity levels are typically an order of magnitude greater in epitaxially grown N-polar GaN than Ga-polar counterpart.\cite{ptak2001controlled,sumiya2000dependence}


\begin{figure*}[t!]
\centering
\includegraphics*[width=11.5cm]{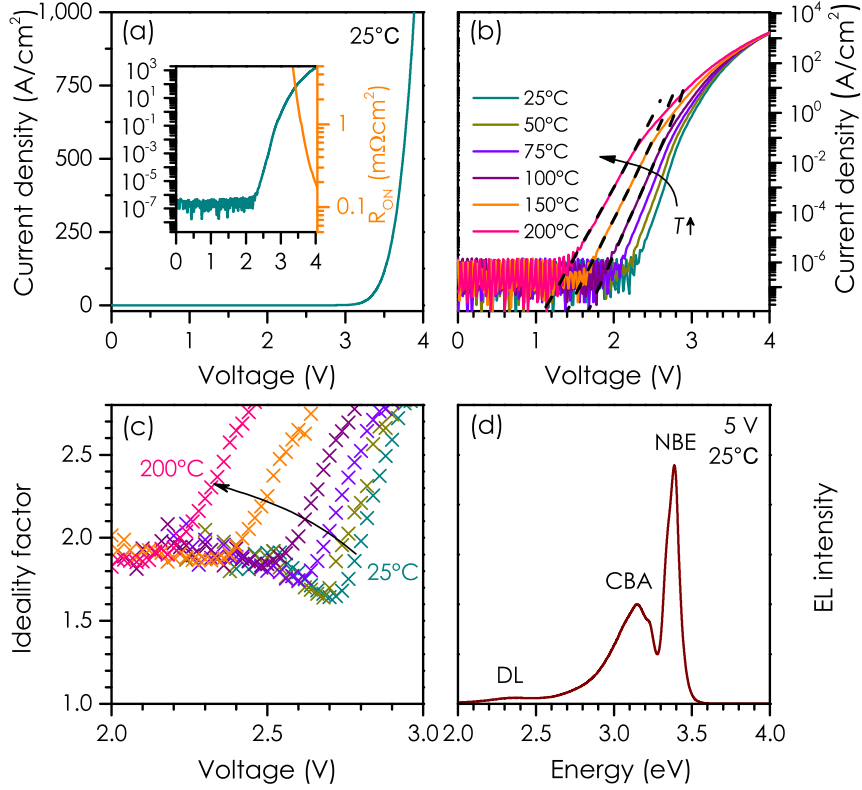} 
\caption{Current density versus voltage (\emph{J-V}) characteristics of the N-polar GaN \emph{p-n} diodes at room temperature in linear scale (a) and in semilog scale [the inset of (a)]. Differential specific resistance (orange line) is also shown in the inset. Temperature dependent \emph{J-V} characteristics (b) and diode ideality factor $\eta$ (c) of the N-polar GaN \emph{p-n} diode. The dashed lines in (b) are calculated Shockley-Read-Hall currents. Random spikes in $\eta$ coming from the derivatives of the noise in \emph{J-V}s at low voltage regions were removed in (c). Note that the apparent ideality factors are kept below 2. (d) Room-temperature electroluminescence spectrum of the N-polar GaN \emph{p-n} diodes measured at a forward bias of 5~V, where near band edge (NBE), conduction band to acceptor (CBA), and deep level (DL) emissions are indicated.} 
\label{forward}
\end{figure*}

We now turn to the transport properties of the fabricated N-polar GaN \emph{p-n} diode devices. Figure~\ref{forward}(a) shows the room-temperature current density-voltage (\textit{J}-\textit{V}) characteristics of a N-polar GaN diode with a diameter of $32$~$\mu$m under forward bias. The leakage current density of the diodes remains lower than a noise floor of 10$^{-6}$~A/cm$^{2}$ for biases from 0 to $\sim2$~V and the on/off current ratio at $\pm4$~V is $>10^{11}$, which is limited by the experimental setup. The diode turns on at $\sim3.5$~V and has a differential on-resistance of $\sim0.2$~m$\Omega$cm$^{2}$ at $\sim4$~V [Fig.~\ref{forward}(a) and the inset]. From the semilog scale of the \emph{J-V} [the inset of Fig.~\ref{forward}(a)], one can see that the current starts to deviate from an exponential function form at $\sim2.8$~V, indicating that the series resistance starts to play an unignorable role in the \emph{J-V} curve above these voltages.\cite{sze2007physics} This effect of the series resistance makes it difficult to extract the true ideality factor at high voltages.\cite{hurni2010pn}

As temperature increases, the forward current density increases and turn-on voltage decreases, and the slope of the current, which is directly related to the diode ideality factor, is seen to decrease [Fig.~\ref{forward}(b)]. For high-quality diodes with negligible structural defects causing leakage current, the total current can be expressed as a sum of diffusion, SRH recombination and radiative current, where the diffusion and the radiative current terms have an ideality factor of 1, whereas the SRH recombination current term has an ideality factor of 2.\cite{sze2007physics,hu2015near} Among these, the radiative current term has the lowest contribution to the total current for typical GaN \emph{p-n} diodes.\cite{hu2015near} The experimental diode current can be expressed by the empirical form $J\propto \text{exp}(qV_{d}/\eta k_{B}T)$, where $q$ is the magnitude of the electronic charge, $V_{d}$ is the diode voltage, $k_{B}$ is the Boltzmann constant, $T$ is temperature and the ideality factor $\eta=(q/k_{B}T)(\partial V_{d}/\partial \text{ln}J)$ has a value between 1 and 2.\cite{sze2007physics} Figure~\ref{forward}(c) shows the temperature dependent $\eta$ calculated from the \emph{J-V} curves in Fig.~\ref{forward}(b). It is seen that the apparent $\eta$ shows a temperature dependent minimal value between 1.6 and 2 in the low bias region. As the bias increases, the apparent ideality factor increases steeply due to the series resistance at high current which shadows the true ideality factor in the high current region.\cite{hu2015near,hurni2010pn} Thus, an ideality factor close to 1 is likely to be observed, if the effects of the series resistance is removed.\cite{hurni2010pn} For high temperatures ($>100$~\celsius), the apparent $\eta$ values are seen to be close to 2, as the diffusion current terms are overwhelmed by the impacts of the series resistance, indicating that the currents in these low bias regions are dominated by the SRH recombination currents. The recombination current can be written as\cite{sze2007physics,hu2015near,maeda2019shockley} 



\begin{equation} \label{SRH}
J_{nr} = \frac{\pi k_{B}Tn_{i}}{\tau_{SRH}E_{0}}\text{sinh}(\frac{qV_{d}}{2k_{B}T}),
\end{equation}

where $n_{i}$ is the intrinsic electron density, $\tau_{SRH}$ is the SRH recombination time and $E_{0}$ is the electric field at the plane with the highest recombination rate which can be approximated as\cite{hu2015near,maeda2019shockley}
  
\begin{equation} \label{field}
E_{0}\approx[\frac{N_{d}k_{B}T(2~\text{ln}~N_{d}-2~\text{ln}~n_{i}-qV_{d}/k_{B}T)}{\epsilon}]^{1/2}.
\end{equation}
Here, $N_{d}$ is the donor density in the uid-GaN region and $\epsilon$ is the dielectric constant. With the temperature dependent $n_{i}$ and an assumption of negligible series resistance in the low biases, Eqs.~(\ref{SRH}) and (\ref{field}) are used to fit the \emph{J-V}s for $>100$~\celsius [the dashed lines in Fig.~\ref{forward}(b)], so that $\tau_{SRH}$ of 0.46, 0.44 and 0.32~ns are extracted at 100, 150, and 200~\celsius, respectively. These $\tau_{SRH}$ values are shorter by almost one order of magnitude than those of state-of-the-art Ga-polar GaN \emph{p-n} diodes,\cite{zongyang} which indicates that the density of SRH non-radiative centers is higher and/or the capture cross-section is larger for N-polar GaN diodes. However, these $\tau_{SRH}$ values are still longer by a factor of several than those in \emph{p}-GaN in Ga-polar \emph{p-n} diodes.\cite{maeda2019shockley} It is interesting to note that the $\tau_{SRH}$ values \emph{decrease} with increasing temperature within the temperature range, contrary to state-of-the-art Ga-polar GaN \emph{p-n} diodes, for which $\tau_{SRH}$ increases with a temperature dependence of $T^{1.4}$.\cite{hu2015near} $\tau_{SRH}$ is given by $1/N_{t}v_{th}\sigma$, where $N_{t}$ is the density of midgap recombination centers, $v_{th}$ ($\propto$ $T^{1/2}$) is the thermal velocity of the carriers and $\sigma$ is the capture cross-section. $\sigma$ varies as $T^{m}$ ($m<0$) as at higher temperature the energetic carriers have to approach more closely to the centres in order to be captured.\cite{tyagi1983minority} The decreasing $\tau_{SRH}$ with increasing temperature, therefore, indicates that $\sigma \propto$ $T^{m}$ with $-0.5<m<0$, i.e., much weaker temperature dependence than those of the Ga-polar diodes, for which $m=-1.9$ (Ref.~\onlinecite{hu2015near}) and $-2.75$ (Ref.~\onlinecite{maeda2019shockley}). 
It has been reported that $\sigma$ highly depends on the complexes of the recombination centers,\cite{chichibu2018origins} therefore different temperature dependence of $\sigma$ (and $\tau_{SRH}$) is likely to be due to different structure of the mid-gap recombination centers, the formation of which could be affected by the polarity and the growth method and conditions. Especially, the impact of the polarity on the formation of different vacancy defects has been recently reported.\cite{mirkhosravi2021impact} However, more studies will be necessary to understand this phenomenon. 

Figure~\ref{forward}(d) shows the room-temperature electroluminescence spectrum measured at a forward bias of 5~V. It is seen that the intense near-band-edge-luminescence (NBE) at $\sim3.4$~eV dominates the spectrum. The deep level (DL) signal at 2.2~eV and the conduction-band-to-acceptor (CBA) emission at $\sim3.1$~eV are both related to Mg acceptors in \emph{p}-GaN.\cite{salviati2000deep,cho2017single} These relatively weak Mg-related emission peaks seem to indicate that the \emph{p}-GaN was not excessively doped with Mg, \cite{salviati2000deep,cho2017single} thus not triggering the formation of compensating defects.\cite{reshchikov2014green} Together with the low background electron density in the uid-GaN, therefore, the success of the high-quality N-polar GaN \emph{p-n} diodes with low ideality factors could be due to lower point defect densities at the junction region. 

\begin{figure}[t!]
\centering
\includegraphics*[width=8cm]{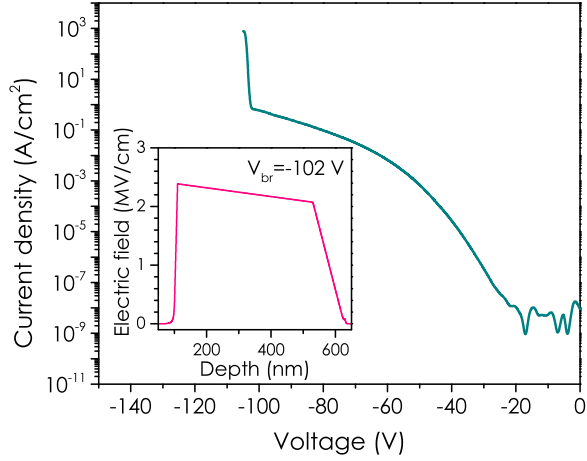} 
\caption{Semilog plot of reverse-bias current density versus reverse-bias voltage characteristics until breakdown for the N-polar GaN \emph{p-n} diodes at room temperature. The inset is simulated electric field profile along the vertical direction of the \emph{p-n}
junction calculated with the breakdown voltage and the doping density. The peak electric field of 2.4~MV/cm is located at the junction.}
\label{reverse}
\end{figure}

We now investigate the breakdown behavior of the N-polar GaN \emph{p-n} diodes. Figure~\ref{reverse}(a) shows the reverse-bias \emph{J-V} characteristics of a 32~$\mu$m diameter N-polar diode without field-plates. With increasing reverse bias, the current density gradually increases from $<10^{-8}$~A/cm$^{2}$ (limited by the experimental setup) to $\sim1~$A/cm$^{2}$ and abrupt electrical breakdown occurs at V$_{br}=$–102~V. Using the uid-GaN
doping density, we calculate the electric field profile at the
breakdown voltage by solving the Poisson equation [the inset of Fig.~\ref{reverse}]. Since the \emph{p}-GaN is more heavily doped than the uid-GaN, the depletion region is
located in the \emph{n}-side of the diode, while the uid-GaN
region is completely depleted, as can be seen in the inset of Fig.~\ref{reverse}. At this breakdown voltage, a peak electric field of 2.4~MV/cm at the edge of the depletion region is estimated, which is still lower than the best Ga-polar GaN \emph{p-n} diodes of $\sim$3~MV/cm,\cite{qi2015high} but higher than the result of a previous report\cite{cho2017single}. The true breakdown behaviors can be unveiled with the use of field-plates in the future, as was done with Ga-polar diodes.\cite{hu2015near} 



In summary, single-crystal N-polar GaN \emph{p-n} diodes with ideality factors less than 2 have been demonstrated by plasma-assisted MBE. The temperature dependent \emph{J-V} characteristics are well explained with the combination of SRH recombination current and diffusion current. At high enough temperatures, where ideality factors are close to 2, SRH recombination times of 0.32--0.46~ns are estimated by experimental data to model fittings. The measured electroluminescence spectrum is dominated by strong near-band edge emission, while deep level and acceptor-related luminescence are greatly suppressed. A relatively high reverse breakdown field of 2.4~MV/cm without field-plates has been achieved. This work indicates that the quality of N-polar GaN diodes can soon be comparable to the state-of-the-art Ga-polar diodes, opening up possibilities for new other high-quality N-polar devices.
\newline


The authors are indebted to Zongyang~Hu and Shyam~Bharadwaj for helping us with the characterization, Takuya~Maeda for useful discussion, Ryan~Page for critical reading of the manuscript, and Vladimir~Protasenko for the maintenance of the experimental apparatus. This work was in part supported by AFOSR (No.~FA9550-20-1-0148). This work made use of the shared facilities that are supported through NSF DMR-1719875 and NSF NNCI-2025233. 


\bibliographystyle{aipnum4-1}
\bibliography{references}
\end{document}